# Muon Nuclear Data

Megumi NIIKURA[1*], Shinichiro ABE[2], Shoichiro KAWASE[3], Teiichiro MATSUZAKI[1],
Futoshi MINATO[4], Rurie MIZUNO[5], Yukinobu WATANABE[3], Yuji YAMAGUCHI[2]

[1]RIKEN Nishina Center, 2-1 Hirosawa, Wako-shi, Saitama 351-0198, Japan

[2]Japan Atomic Energy Agency,
2-4 Shirakata, Tokai-mura, Naka-gun, Ibaraki 319-1195, Japan

[3]Faculty of Engineering Sciences, Kyushu University
6-1 Kasugakoen, Kasuga-shi, Fukuoka 816-8580, Japan

[4]Faculty of Science, Kyushu University, 744 Motooka, Nishi-ku, Fukuoka 819-0395, Japan

[5]Faculty of Science, the University of Tokyo, 7-3-1 Hongo, Bunkyo-ku, Tokyo 113-0033, Japan

[*]Email: niikura@ribf.riken.jp

We plan to develop a new nuclear database for muon-induced nuclear reactions (muon nuclear data). The database will consist of (1) energies and intensities of the muonic X rays, (2) lifetimes of the muonic atom, (3) production branching ratio of the residual nuclei by muon capture, (4) emission probabilities of the particles after muon capture, and (5) energy spectra of the emitted particles after muon capture. In this paper, we review the present status and current investigations for the muon nuclear data.

## 1. Introduction

The muon is a second-generation charged lepton with a mass of 105.658 MeV/c$^2$ and a lifetime of approximately 2.2 µs in a vacuum. When a negatively charged muon[1] stops in matter, it forms a highly-excited atomic state with the nucleus, known as a muonic atom. The excited muonic atom promptly cascades down to the atomic ground (1s) state by emitting Auger electrons and muonic X rays. Because the muon's wave function in the muonic atom largely overlaps with that of the nucleus, the energies of the muonic X rays are affected by the finite size of the nuclear charge distribution and thus provide a determination of an absolute charge radius of the nucleus [1–4]. The energies of the muonic K$_\alpha$ X rays for various nuclei are summarized in Ref. [1] as a table of nuclear charge radii.

The muon in the muonic atom at the 1s state decays via two processes: free muon decay and muon capture. The free muon decay is the same decay process as the muon in a vacuum as follows:
$$\mu^- \to e^- + \nu_\mu + \nu_e, \quad (1)$$
and muon capture is a process which is the capture of the muon by a proton in a nuclear medium via the charged current of the weak interaction as follows [5]:
$$\mu^- + p \to n + \nu_\mu. \quad (2)$$
The partial lifetime of the free muon decay is approximately the same as the lifetime of the negative muon in a vacuum, and that of muon capture depends on the overlap of the wave functions of the muon and the nucleus. Because the size of the muonic atom (Bohr radius) is approximately inversely proportional to the element number ($Z$) of the nucleus, the lifetime of the muonic atom is shorter for the heavy muonic atom. The typical lifetime of the muonic atom with $Z < 10$ is 2 μs, corresponding to the free decay branch of more than 90%. In contrast, the typical lifetime for the heavier muonic atom with $Z > 40$ is about 100 ns, corresponding to the free decay branch of less than 5% (see Fig.1 in Ref. [5]). Most of the published data for measured lifetimes of the muonic atom are listed in Ref. [6].

---

[1] Though there are two muons, the positive and negative muons, we discuss only the negative muon in this paper, and hereafter "muon" represents the negatively charged muon.

Muon capture is an analogous reaction to electron capture; the main difference between them is in their energy transfer. The energy released by muon capture is 104.3 MeV, corresponding primarily to the mass of the muon. If the proton is at rest, as expressed in Equation (2), the recoiling neutron takes only 5.2 MeV of kinetic energy, whereas the neutrino takes away 99.1 MeV. Muon capture for the general nucleus of $(A, Z)$, where $A$ is the mass number of the nucleus, produces a compound nucleus of $(A, Z-1)^*$ as follows:

$$\mu^- + (A,Z) \rightarrow (A, Z-1)^* + \nu_\mu. \qquad (3)$$

Because the nucleus is a many-body system, the excitation energy of the compound nucleus is distributed around 10–50 MeV. The energy and angular-momentum (spin) distributions of the compound states populated by muon capture are of interest in understanding of the reaction mechanism; however, experimental data are scarce and require improvements.

The importance of the interactions between the muon and the nucleus, namely spectroscopic information of the muonic atom and muon capture, has recently been focused on in the many fields of the natural sciences and applications, such as nuclear physics, nuclear transmutation for nuclear waste [7], muon-induced radioactive isotope production for medical use [8], radiation safety data in the muon facilities, cosmic muon-induced soft error in modern semiconductor devices [9,10], cosmogenic production of radioactive nuclides for geological studies, and non-destructive element analysis by muon-induced X-ray emission (MIXE) [11-13]. Despite those demands, nuclear data for muon-induced nuclear reactions (muon nuclear data) has not been established thus far.

## 2. Muon nuclear data

We here propose to develop muon nuclear data. The database will consist of the following five sublibraries:
(1) The energies and intensities of the muonic X rays,
(2) Lifetimes of the muonic atom (muon capture probability),
(3) The production branching ratio of the residual nuclei by muon capture,
(4) Emission probabilities and multiplicities of the evaporation particles by muon capture,
(5) Energy distribution of the emission particles by muon capture.

Important note that the "cross-section" is not an objective of muon nuclear data because the formation probability of the muonic atom is unity, and the reactions occur only when the muon stops in the matter.

## 3. Current status of muon nuclear data

The muonic X-ray energies only for the Lyman series (Kα) are summarized in Ref. [1], and measured X-ray energy spectra for various samples are uploaded on the website by the Joint Institute for Nuclear Research (JINR) [14]; however, energies for other higher-order X-ray series and muonic X-ray intensities are rarely found in the literature. For the theoretical model calculation, the muonic X-ray energies can be calculated by Mudirac [15]. Mudirac reproduced measured muonic X-ray energy rather well, within approximately 1-keV accuracy, while the calculation for the X-ray intensity is under development.

Most of the published data for measured lifetimes of the muonic atom are listed in Ref. [6]. For the theoretical model calculation, Primakoff proposed an empirical formula to estimate the lifetimes of the muonic atom [16]. The Primakoff formula well reproduced experimentally measured lifetimes with some exceptions (see Fig. 6 in Ref. [6]).

The production branching ratios of the residual nuclei by muon capture have been measured in the past, and most of them are summarized in the review article by Measday [17]. The production branching ratios of reaction residues of muon capture were estimated from the known experimental data, for example, as shown in Table 5.5 in Ref. [17] for a case of muon capture on $^{28}$Si. Because of several inconsistencies in the different experiments, Measday "gives no error because of the inconsistencies" and states this compilation as a "suggestion".

The production branching ratio of the residual nuclei and emission probabilities and multiplicities of the evaporation particles by muon capture are complementary information. In medium and heavy nuclei, the particles emitted from muon capture are primarily neutrons because the emission of charged particles is suppressed by the Coulomb barrier. Neutron multiplicity has been measured in the past using a large liquid scintillator tank [18], γ-ray measurement [19], and activation measurement [20]. The neutron multiplicities for 0, 1, 2, and 3 neutron emissions have been estimated at 15-30%, 50-60%, 10-20%, and 0.2-10%, respectively [20]. Because of the large error propagation in the folding analysis for the neutron measurement and the incomplete data for the activation, only roughly estimated values without uncertainty for the neutron multiplicity are known so far.

The energy distribution of the emission particles (neutron, proton, deuteron, alpha, etc…) by muon capture has also been measured for some nuclei. The energy spectra of neutrons have been measured for the heavy nuclei of Tl, Pb, Bi [21], and Pd isotopes [22]. The low-energy component of the neutron energy spectrum below 5 MeV can be understood by the statistical evaporation from the compound nucleus; however, the spectrum extends to higher energies. The high-energy component of the neutron energy spectrum is interpreted as due to direct and pre-equilibrium processes, in which the neutron is emitted immediately at the time of muon capture before reaching the thermal equilibrium of the compound states. The energy spectra for the charged particles were also measured for some nuclei; however, the experimental data is limited.

## 4. Recent investigations for muon nuclear data

In this section, we introduce our recent theoretical and experimental investigations and developments related to muon nuclear data.

### 4.1. Theoretical models

For the theoretical model calculation, muon interaction models have recently been implemented in the Monte Carlo simulation code of the particle and heavy ions transport code system (PHITS) [23,24]. In this model, the neutron energy produced by muon capture, as expressed in Equations (2) and (3), was sampled from the excitation function proposed by Singer [25]. The time evolution of the initial neutron energy to the compound nucleus was calculated using JAERI Quantum Molecular Dynamics (JQMD) [26,27], and the sequential evaporation process was calculated using the Generalized Evaporation Model (GEM) [28].

There is another attempt to calculate the muon capture reaction by the microscopic theoretical model [29,30]. In the model, the muon capture rates are estimated by the second Tamm-Dancoff approximation with the two-component exciton model, which describes particle emission from the pre-equilibrium state [31,32]. For particle evaporation from the compound state, the Hauser-Feshbach statistical models were applied [33].

Due to the lack of muon nuclear data, particularly related to the muon capture reaction, the model calculation requires further experimental data for benchmarking and improvements.

### 4.2. Developments of experimental methods and detectors

The activation method is the most reliable and sensitive technique for determining the production rate of radioactive nuclei by muon beam irradiation. In classical activation measurements, only production ratios of long-lived radioactive isotopes can be obtained because the decay measurements usually take place separately at the time and location of beam irradiation to avoid the beam background. We have recently developed a novel method called the in-beam activation method, which enables us to measure the activities of short-lived states [34]. The combination of in-beam and ordinary offline activation methods enables the measurement of most of the β-decaying states with a wide range of half-lives from a few milliseconds to years.

Recently, we have also developed detectors for muonic X-ray spectroscopy [35,36] and for the measurements of neutrons and charged particles from muon capture [37,38]. The wide-range photon detection system for muonic X-ray spectroscopy was developed using a germanium detector with Compton suppressors. The system was demonstrated at the muon facility in the Paul Scherrer Institute, and complete data of muonic X-ray energies and intensities for $^{197}$Au and $^{209}$Bi was obtained [38]. For the neutron measurement, a new solid-state scintillator with neutron-gamma discrimination capability (EJ-276) is under development [37]. The charged particle detector array was also constructed. The detector has a particle discrimination capability based on digital waveform analysis techniques for neutron transmutation-doped silicon (nTD-Si) detectors [38]. In 2023, charged particle measurement from muon capture of $^{28}$Si at the RIKEN-RAL facility in the Rutherford Appleton Laboratory was performed, the data analysis of which is ongoing.

## 5. Summary and outlook

We plan to develop the muon nuclear data, which will consist of the five sublibraries. The importance of the interactions between the muons and the nuclei, namely the spectroscopic information of the muonic atom and muon capture, has recently been focused on in the many fields of the natural sciences and applications. Although some of the sublibraries measured in the past are already summarized in the publications [1,6], there is no complete database, and it needs compilation and improvement. Several theoretical and experimental investigations for muon nuclear data are currently ongoing.